\documentstyle[12pt]{article}
\begin{document}
\baselineskip=2\baselineskip
\begin{center}
{\large\bf  Theory on the Temperature Dependence }  \\
\end{center}
\begin{center}
{\large\bf  of Giant Magnetoresistance}
\end{center}
\begin{center}
Hideo Hasegawa\footnote{E-mail address: hasegawa@u-gakugei.ac.jp}
\end{center}
\begin{center}
{\it Department of Physics, Tokyo Gakugei University  \\
Koganei, Tokyo 184, Japan}
\end{center}
\thispagestyle{myheadings}
%
\begin{center}
{\bf Abstract}   \par
\end{center}

    The temperature dependence of the giant magnetoresistance (GMR)
for currents parallel and perpendicular to the multilayer plane,
is discussed by taking account of the random exchange potentials,
phonon scatterings and spin fluctuations.  The effect of spin
fluctuations, which plays an important role at finite temperatures,
is included by means of the static functional-integral method developed
previously by the present author. Our model calculations well
explain the observed features of the parallel and perpendicular
GMR of Fe/Cr and Co/Cu multilayers recently reported by Gijs {\it et al}.

\vspace{3.0cm}
\noindent
PACS Numbers: 73.50.Bk,  73.50.Jt,  75.50.Rr

\vspace{1.0cm}
\noindent

\newpage
\noindent
{\large\bf I. INTRODUCTION}

   The giant magnetoresistance (GMR)$^{1}$ in magnetic multilayers is
one of the most attractive subject in current solid-state physics.
In recent years much progress has been made in understanding the
GMR and its related phenomena.$^{1}$
One of the important aspects of the GMR is its temperature
dependence.  A careful study of the temperature dependence
of GMR is not only important in understanding its mechanism
but also very useful to its realistic applications.
Most of the magnetic multilayers are fabricated with
transition metals such as Fe, Ni and Co.
It would be instructive to briefly discuss
the temperature dependence of
the resistivity of $bulk$ transition metals,
before we study the temperature dependence of the GMR or
of the resistivity of transition-metal multilayers.
It has been reported that when the temperature is raised from T = 0 K,
the resistivity of Fe$\:\:^2$ or Co$\:\:^3$ gradually increases up to the
Curie temperature, where it has a cusp (see Fig.1).
This characteristic
temperature dependence of the resistivity is
interpreted as due to the contributions
from impurity, phonon and magnetic terms.
The last contribution is classically discussed
as spin-disorder scatterings
with the use of the s-d model.$^4$  Lately, a modern theory
on the itinerant-electron magnetism has accounted for it
in terms of spin fluctuations.$^5$

  It has been well known that d-electrons in transition metals
show both the localized and itinerant character:
the Curie-Weiss susceptibility and the large specific heat peak near
the Curie temperature are easily explained by
the localized-spin model whereas
the non-integral ground-state moment and the large
linear-specific heat coefficient favor the band model.
It has been realized that the effect of spin fluctuations plays
essential  roles to reconcile the duality of d electrons.$^6$
The finite-temperature band theory, which  has been
proposed by Hasegawa,$^7$  includes the effect
of spin fluctuations by means of the static functional-integral
method combined with the coherent potential approximation (CPA).
Spin fluctuations including spin waves are shown to yield
the $T^2$ contribution to the resistivity at $T \simeq 0$
by several approaches.$^{8}$
This type of theories$^8$ is, however, valid
only at very low temperatures.
In our finite-temperature theory,$^7$ spin fluctuations are
regarded as localized, static modes with the adopted approximations.
This method has proved useful in understanding the overall
finite-temperature properties of transition metals, alloy and
multilayers,$^{9}$
covering both below and above the Curie temperature.

By employing the finite-temperature band theory,$^7$
we discussed in previous papers$^{10}$ the temperature
dependence of the MR ratio for
currents parallel to the multilayer plane.
The observed temperature dependences of Fe/Cr,$^{10d,11}$
NiCo/Cu, NiFe/Cu and CoFe/Cu$^{12}$ multilayers
have been shown to be well explained by our theory.
It has been pointed out$^{9c}$ that a multilayer in which the normal
and {\it inverse} GMR$^{13}$ coexist, may have an interesting
temperature dependence beneficial for real applications.

One of the purposes of the present paper is to generalize
our theory$^{10}$ to the perpendicular GMR,
whose experimental$^{14-16}$ and theoretical study$^{17-20}$
has been currently performed.
The other purpose is to include the phonon contribution to
the conductivity calculation, which was neglected
in our previous study.$^{10}$
The paper is organized as follows:
In the Sec.II, we present our formulation applying
our finite-temperature band theory to the GMR.
Numerical calculations of the parallel and perpendicular GMR
of Fe/Cr and Co/Cu multilayers
are reported in Sec.III.
Supplementary discussions are given in Sec.IV.
\vspace{1.0cm}
\noindent
{\large\bf II. CALCULATION METHOD}

\noindent
A. {\it An Adopted Model and the Expression of GMR }

   We adopt an A/B multilayer consisting of magnetic A and
nonmagnetic B atoms
with the simple-cubic (001) interface.
The layer parallel to the interface is assigned by the index
$n \: (=1 - N_f)$.
The thickness of the A and B layers is assumed to be thinner
than the mean free path  and  sufficiently thin
compared with the spin diffusion length.$^{17}$
It is assumed that atoms  A  and B  are
randomly distributed on layer $n$
with the concentrations of
$x_n$ and $y_n$,
respectively
$ \: (x_n + y_n = 1)$.
The film is described by the
single-band Hubbard model,
in which the atomic potential
(the on-site interaction)
is  assumed to be given by
$\varepsilon^A$ and $\varepsilon^B$  ($U^A$ and $U^B$)
when a given lattice site is occupied by A and B atoms,
respectively.

    In order to study the finite-temperature properties of
the magnetic film,
we apply the functional-integral method within the
static approximation to the Hubbard Hamiltonian.$^7$
The partition function is evaluated by calculating
the partition function of the effective one-electron
Hamiltonian including  the random charge and exchange fields with
the Gaussian weight.  The charge field is include by
the saddle-point approximation and the exchange field
by the alloy-analogy approximation with the CPA.
The energy-dependent coherent potential for an
$s$-spin electron ($s = \uparrow, \downarrow$)
on the layer $n$, $\Sigma_{ns}(\varepsilon)$,
is determined by the CPA condition.
The coherent potentials, the average of the magnetic moments
on the layer $n$, $\langle M_n \rangle$,
and its root-mean-square (RMS) value,
$\langle (M_n)^2 \rangle^{1/2}$,
are calculated self-consistently, details having been given
in Ref.7.

%


  When we employ the CPA, the conductivity of the film
is given by$^{20}$
\begin{equation}
\sigma_{\xi \eta} = \left(\frac{e}{\hbar} \right)^2
\left( \frac{1}{\pi} \right)  \int d\varepsilon
\left( -\frac{\partial f}{\partial \varepsilon} \right) \:
{\rm Tr} \left(v_{\xi}
\: {\rm Im} {\bf G} \: v_{\eta}
\: {\rm Im} {\bf G} \right)
\:\:\:\:\:\:  (\xi, \eta = x, y, z),
\end{equation}
provided the vertex correction is neglected.
In Eq.(1) $v_{\xi}$ is the velocity operator
and ${\bf G}$ is the Green function matrix.
The  conductivities
for currents parallel ($\parallel$) and
perpendicular ($\perp$) to the film layer
are  given by$^{10,18,20}$
\begin{equation}
\sigma^{\parallel} =
\left(\frac{e}{\hbar}\right)^2
\: \int d\varepsilon \:
\left( -\frac{\partial f}{\partial \varepsilon} \right) \:
\sum_s \nu^{\parallel}(\varepsilon)
\left(\frac{1}{N_f} \right) \sum_n \sum_m
\frac{a_{nms} \: \tau_{nms}}{(\Delta_{ns} + \Delta_{ms})},
\end{equation}
\begin{equation}
\sigma^{\perp} =
\left(\frac{e}{\hbar}\right)^2
\: \int d\varepsilon
\left( -\frac{\partial f}{\partial \varepsilon} \right) \:
\: \sum_s \nu^{\perp}(\varepsilon)
\left[ \left( \frac{1}{N_f} \right)
\sum_n \Delta_{ns} \right]^{-1},
\end{equation}
with
\begin{equation}
\nu^{\lambda}(\varepsilon) = \hbar^2 \sum_{k_{\parallel}} v_{\lambda}^2  \:\:
\delta(\varepsilon - \varepsilon_{k_{\parallel}})   \:\:\:\:\:\:
(\lambda = \parallel, \perp),
\end{equation}
\begin{equation}
\tau_{nms}= \delta_{nm} + (1 - \delta_{nm})
  \left( \frac{(\Delta_{ns} + \Delta_{ms})^2}
 {\left[(\Lambda_{ns}-\Lambda_{ms})^2
+(\Delta_{ns} + \Delta_{ms})^2  \right]} \right),
\end{equation}
which is valid within the Born approximation.
In Eqs. (2)-(5) $\Lambda_{ns}$ = Re $\Sigma_{ns}(\varepsilon)$,
$\Delta_{ns} = \: \mid\:$Im $\Sigma_{ns}(\varepsilon)\mid $,
$\Sigma_{ns}$ is the coherent potential of an $s$-spin
electron on layer $n$, and
$a_{nls}$ and $\nu^{\lambda}$ are specified
by the electronic structure of the  film
(see Eqs. (19) and (20) in Ref. [10a]).
Analytic expressions given by Eqs.(2)-(5) have clear physical
meaning. When currents flow parallel to the plane, an $s$-spin electron
propagating successively from a site on layer $n$ to a site
on layer $m$, is scattered with the strength proportional to
$\Delta_{ns}$ and $\Delta_{ms}$, respectively, and
its conductivity is given as a sum of such processes
with the weight of $a_{nms}\tau_{nms}$.$^{10}$
On the contrary, in the case of the  perpendicular current,
the $s$-spin conductivity is  given as of a series circuit
of resistivities on successive layers,
each of which is  proportional to $\Delta_{ns}$.$^{18,21}$
In both cases, the total conductivity is a sum of
the up- and down-spin channels.
The so-called spin-flop process is implicitly included
through the spin-fluctuation term
which is responsible to a decrease in layer magnetization,
as will be shown shortly.
In the next section, we will employ our formalism
in a semi-phenomenological way
to discuss the temperature
dependence of the MR ratio.

\vspace{1.0cm}
\noindent
B. {\it A Semi-phenomenological Study of GMR}

   We adopt a system  consisting of
magnetic $({\rm M}_1, {\rm M}_2)$
and nonmagnetic $({\rm N}_1, {\rm N}_2)$ layers,
whose  thickness
are $M$ and $N$, respectively.
Bulk scatterings are assumed to be important in these layers,
related discussion will be given in Sec.IV.
When magnetic moments on ${\rm M}_1$ and ${\rm M}_2$ layers are
in the antiferromagnetic
(AF) configuration,
the real and imaginary parts of the coherent potentials
are given by$^{10}$
\begin{eqnarray}
\Lambda_{ns}^{\rm AF} - i \Delta_{ns}^{\rm AF}
&=& \Lambda_s - i \Delta_s \qquad \qquad
\mbox{for $ n \in {\rm M}_1, $}  \\
&=& \Lambda_{-s} -i \Delta_{-s} \quad \qquad
\mbox{for $ n \in {\rm M}_2,$} \\
&=& \Lambda_0 - i \Delta_0 \qquad \qquad
\mbox{for $ n \in {\rm N}_1, {\rm N}_2.$}
\end{eqnarray}
\noindent
Using  Eqs.(2)-(8), we get the parallel and perpendicular
conductivities given by$^{10}$
\begin{equation}
\sigma^{{\rm AF} \parallel}
= \sum_s \left\{ \frac{2c_{\rm MM}^{\rm AF}}{(\Delta_s + \Delta_{-s})}
+ \frac{c_{\rm NN}^{\rm AF}}{\Delta_0}
+ 4c_{\rm MN}^{\rm AF} \left( \frac{1}{\Delta_{s} + \Delta_0}
            +  \frac{1}{\Delta_{-s}+\Delta_0} \right)
+ d_{\rm M}^{\rm AF}\left(\frac{1}{2\Delta_{s}}
            + \frac{1}{2\Delta_{-s}} \right)
+ \frac{d_{\rm N}^{\rm AF}}{\Delta_0} \right\} ,
\end{equation}
\begin{equation}
\sigma^{{\rm AF} \perp}
=\left( \frac{e}{\hbar} \right)^2 \nu^{\perp} N_f
\sum_s \left\{ \frac{1}{M \Delta_s + M \Delta_{-s} + 2N \Delta_0}  \right\},
\end{equation}
\noindent
with
\begin{equation}
c_{\rm MM}^{\rm AF} = N_f^{-1} (e/h)^2  \nu^{\parallel}
\sum_{n \in {\rm M}_1}  \sum_{m \in {\rm M}_2} a_{nm} \tau_{nm},
\end{equation}
\begin{equation}
d_{\rm M}^{\rm AF} = N_f^{-1}  (e/h)^2  \nu^{\parallel}
\sum_{n \in {\rm M}_1}  \sum_{m \in {\rm M}_1} a_{nm} \tau_{nm},
\end{equation}
\noindent
and $c_{\rm NN}^{\rm AF}$, $c_{\rm MN}^{\rm AF}$,
and $d_{\rm N}^{\rm AF}$
are given by similar expressions.
We employed the $T=0$ limit of Eqs. (2) and (3)
because the relevant temperature is much less than
the Fermi energy, $\varepsilon_{\rm F}$.
In Eqs.(9)-(12) $\nu^{\lambda} = \nu^{\lambda}(\varepsilon_{\rm F})$,
$N_f = 2(M+N)$, and
the spin dependence
in $a_{nms}$ and $\tau_{nms}$ is neglected.
Subscripts, MM, NN and MN,
denote the contributions
from the $interlayer$ scatterings between magnetic layers,
between nonmagnetic layers, and between magnetic and nonmagnetic
layers, respectively.  On the contrary, the single subscript, M (N),
expresses the contribution from the $intralayer$
scatterings within magnetic (nonmagnetic)
layers.

    On the contrary, when magnetic moments
on the subsequent magnetic layers are
in the ferromagnetic (F) configuration,
the real and imaginary parts of the
coherent potentials are given by$^{10}$
\begin{eqnarray}
\Lambda_{ns}^{\rm F} - i \Delta_{ns}^{\rm F}
&=& \Lambda_s - i \Delta_s \qquad \qquad
\mbox{for $ n \in {\rm M}_1, {\rm M}_2$},  \\
&=& \Lambda_0 - i \Delta_0 \qquad \qquad
\mbox{for $ n \in {\rm N}_1, {\rm N}_2$}.
\end{eqnarray}
\noindent
We get the parallel and perpendicular conductivities given by$^{10}$
\begin{equation}
\sigma^{{\rm F} \parallel}
= \sum_s \left\{ \frac{c_{\rm MM}^{\rm F}}{\Delta_s}
+ \frac{c_{\rm NN}^{\rm F}}{\Delta_0}
+ \frac{8c_{\rm MN}^{\rm F}}{(\Delta_s + \Delta_0)}
+ \frac{d_{\rm M}^{\rm F}}{\Delta_s}
+ \frac{d_{\rm N}^{\rm F}}{\Delta_0} \right\},
\end{equation}
\begin{equation}
\sigma^{{\rm F} \perp}
=\left( \frac{e}{\hbar} \right)^2 \nu^{\perp} N_f
\sum_s \left\{ \frac{1}{2M \Delta_s + 2N \Delta_0}  \right\}.
\end{equation}

The MR ratio,
$\Delta R/R$,
is given
from Eqs. (9), (10), (15) and (16), by
\begin{equation}
\left( \frac{\Delta R}{R} \right)^{\lambda}
\equiv  \frac{(R^{\rm AF} - R^{\rm F})}{R^{\rm F}}
= \frac{(\sigma^{\rm F} - \sigma^{\rm AF})}{\sigma^{\rm AF}}
= \frac{(\alpha - \beta)^2}
{4 \alpha \beta} X^{\lambda}
\:\:\:\:\:\:  (\lambda = \parallel, \perp),
\end{equation}
with
\begin{equation}
\alpha = \Delta_{\uparrow}/\Delta_0,    \:\:\:\:
\beta = \Delta_{\downarrow}/\Delta_0,
\end{equation}
\begin{equation}
X^{\parallel}  = \left[ 1 + g_0 \frac{(\alpha+\beta)^2}{\alpha \beta}
+ g_1\left(\frac NM\right)(\alpha+\beta)
\left( \frac {1}{\alpha+1} + \frac {1}{\beta+1} \right)
+ g_2 \left( \frac NM \right)^2 (\alpha + \beta) \right]^{-1},
\end{equation}
\begin{equation}
X^{\perp} = \left[1 +
\left( \frac{N}{M} \right) \left( \frac{1}{\alpha} + \frac{1}{\beta} \right)
 + \left( \frac{N}{M} \right)^2 \left( \frac{1}{\alpha \beta}\right)
\right]^{-1}.
\end{equation}
In Eq. (19) $g_0$, $g_1$ and $g_2$ are defined by$^{10}$
\begin{equation}
\frac{d_M}{4c_{\rm MM}} = g_0, \quad
\frac{2c_{\rm MN}}{c_{\rm MM}} = g_1 \left(\frac{N}{M}\right), \quad
\left( \frac{c_{\rm NN}+d_N}{2c_{\rm MM}}\right)
= g_2 \left(\frac{N}{M}\right)^2.
\end{equation}
The expression for the GMR given by Eqs.(17) and (20)
is just the same as that derived by Edwards {\it et al.}$^{21}$
using the resistor network model and
has been employed for an analysis of $(\Delta R/R)^{\perp}$.$^{14}$

   Setting $N = 0$ in Eqs. (17)-(20), we get
\begin{equation}
\left( \frac{\Delta R}{R} \right)^{\parallel}
= \frac{(\alpha - \beta)^2}   {4 \alpha \beta
\left[ 1 + g_0 (\alpha + \beta)^2/\alpha \beta \right]},
\end{equation}
\begin{equation}
\left( \frac{\Delta R}{R} \right)^{\perp}
= \frac{(\alpha - \beta)^2}   {4 \alpha \beta},
\end{equation}
and the ratio of the parallel GMR to the perpendicular one is given by
\begin{equation}
\frac{\left( \Delta R/R \right)^{\parallel}}
{\left( \Delta R/R \right)^{\perp}} = \frac{1}{[1 + g_0 (a + 1)^2/a]}
\: \leq \: 1 \:\:\:\:\:\:\:\: (a = \alpha/\beta).
\end{equation}
When electrons flow perpendicular to the layer plane, all electrons
pass through the adjacent two magnetic layers.
On the contrary, it is not the case
for currents parallel to the plane; some electrons go through
only the one of magnetic layers without probing the other magnetic layer.
The second $g_0$ term of the denominator of Eq.(22) denotes this contribution.
Itoh {\it et al.}$^{20}$ claim that the anisotropy of the velocity
operator : $\nu^{\parallel}/\nu^{\perp} \geq 1 $
is the main mechanism leading to
$(\Delta R/R)^{\parallel} \leq (\Delta R/R)^{\perp}$.
The factor, $\nu^{\parallel}$ or $\nu^{\perp}$, is not,
however, relevant because it is
{\it cancelled out} when the MR ratio given by
Eq.(17) is calculated.

   The temperature dependence of the GMR arises from
those of $\alpha$ and $\beta$, which is expressed in terms of
the coherent potential of the film (Eq.(18)),
whose imaginary part
in the magnetic (${\rm M}_1$ or ${\rm M}_2$) layer
is given within the Born approximation by$^{10}$
\begin{equation}
\Delta_s = \Delta_s^r + \Delta_s^s + \Delta_s^p,
\end{equation}
with
\begin{equation}
\Delta_s^r = \pi \rho_s x \: y \:
[\tilde{\varepsilon}^{A} - \tilde{\varepsilon}^{B}
-  s \left( \frac{U^{A}}{2} \right) \langle M^{A} \rangle ]^2,
\end{equation}
\begin{equation}
\Delta_s^s = \pi \rho_s \: x \:
\left( \frac{U^{A}}{2} \right)^2
[\langle (M^A)^2 \rangle - \langle M^A \rangle^2] ,
\end{equation}
\begin{equation}
\Delta_s^p = P_m \: \rho_s Z(T/\Theta_m),
\end{equation}
where
$\tilde{\varepsilon}^{A}$ and
$\tilde{\varepsilon}^{B}$ are
the spin-independent Hartree-Fock potentials
and $\rho_s$ is the density of states of an $s$-spin electron
at the Fermi level.
The first term ($\Delta_s^r$) in Eq. (25) arises from
the scattering due to random Hartree-Fock potentials
for an $s$-spin electron:
the second term ($\Delta_s^s$) comes from the effect of
spin fluctuations:
the third term ($\Delta_s^p$) is introduced for phonon scatterings whose
explicit form will be given shortly (Eq.(45)).

   On the other hand,
the imaginary part of the coherent potential
in the nonmagnetic
(${\rm N_1}$ or ${\rm N_2}$) layer, is given by
\begin{equation}
\Delta_0 = \Delta_0^r + P_n \: \rho_0 Z(T/\Theta_n),
\end{equation}
where the first and second terms denote
the contributions from
random potentials and phonons, respectively, and
$\rho_0$ is the density of states at the Fermi level
of the nonmagnetic metal.
In Eqs.(28) and (29) $\Theta_m$ and
$\Theta_n$ are Debye temperatures, and
$P_m$ and $P_n$ are related with the
electron-phonon interactions in magnetic and nonmagnetic metals.

   Using Eqs. (18), (26)-(29),
we get $\alpha$ and $\beta$ given by

\begin{equation}
\alpha
= A  \frac{ (1 + \gamma(T))  \left[ x y (B + m(T))^2
  + x (\mu(T)^2 - m(T)^2) + p_m Z(T/\Theta_m) \right]  }
     {(1 + p_0 Z(T/\Theta_n))},
\end{equation}
\begin{equation}
\beta
= A  \frac{ (1 - \gamma(T))  \left[ x y (B - m(T))^2
  + x (\mu(T)^2 - m(T)^2) + p_m Z(T/\Theta_m) \right]  }
     {(1 + p_0 Z(T/\Theta_n))},
\end{equation}
with
\begin{equation}
m(T) = \langle M^A \rangle /M_0,
\end{equation}
\begin{equation}
\mu(T) = \sqrt{ \langle (M^A)^2 \rangle }/M_0,
\end{equation}
\begin{equation}
\gamma(T) = (\rho_{\uparrow} - \rho_{\downarrow})
/(\rho_{\uparrow} + \rho_{\downarrow} ),
\end{equation}
\begin{equation}
A = \pi \rho (U^A M_0/2)^2 /\Delta_0^r ,
\end{equation}
\begin{equation}
B = (2/U^A M_0)
(\tilde{\varepsilon}_{B} - \tilde{\varepsilon}_A ),
\end{equation}
\begin{equation}
p_{m} = P_{m} /\pi  (U^A M_0/2)^2,
\end{equation}
\begin{equation}
p_0 = P_n \: \rho_0 /\Delta_0^r
= p_m \: A \: (P_n/P_m)\:(\rho_0/\rho),
\end{equation}
where
$\rho=(1/2)(\rho_{\uparrow}+\rho_{\downarrow}) $ and
$M_0$ is the ground-state magnetic moment.

At $T = 0 $K where $m(0) = \mu(0) = 1$ and $\gamma(0) = \gamma_0$,
Eqs. (30) and (31) become
\begin{equation}
\alpha_0 = \alpha(T = 0) = x y \: A \: (1 + \gamma_0)(B + 1)^2, \: \:
\end{equation}
\begin{equation}
\beta_0 = \beta(T = 0) = x y \: A \: (1 - \gamma_0)(B - 1)^2,
\end{equation}
from which the coefficients A and B are expressed
in terms of $\alpha_0$, $\beta_0$ and  $\gamma_0$ as
\begin{equation}
A = \frac{1}{4 x y} \left( \sqrt{\frac{\alpha_0}{1 + \gamma_0}}
   - \sqrt{\frac{\beta_0}{1 - \gamma_0}} \right)^2,
\end{equation}
\begin{equation}
B = \left( \sqrt{\frac{\alpha_0}{1 + \gamma_0}}
+ \sqrt{\frac{\beta_0}{1 - \gamma_0}} \right)
/\left( \sqrt{\frac{\alpha_0}{1 + \gamma_0}}
- \sqrt{\frac{\beta_0}{1 - \gamma_0}} \right),
\end{equation}

    The normalized magnetic moment, $m(T)$,  and its RMS value,
$\mu(T)$, are in principle calculated
with the use of the finite-temperature band theory.$^{9}$  We here, however,
adopt simple, analytic expressions of $m(T)$ and $\mu(T)$
for our model calculation, given by$^{10}$
\begin{equation}
m(T)= \sqrt{1 - (T/T_C)^2},  \:\:\:\:\:
\mu(T) = 1.
\end{equation}
The temperature dependence of the spin asymmetry $\gamma(T)$ defined by
Eq.(34) is assumed to be given by
\begin{equation}
\gamma(T) = \gamma_0 \:\: m(T).
\end{equation}
As for the  phonon contribution given by $Z(T/\Theta_m)$
in Eqs.(28) and (29),
we adopt the simple Gr\"{u}neisen function:
\begin{equation}
Z(T/\Theta_m) =  (T/\Theta_m)^5
\int_0^{\Theta_m/T} dy \: \frac{y^5}{(e^y-1)(1-e^{-y})},
\end{equation}
which is $124.43 \: (T/\theta_m)^5$ at $T/\Theta \ll 1$
and $T/4\Theta_m$   at $T/\Theta_m \gg 1$.

   Now we may  calculate the MR ratio, $\Delta R/R$, as a function of
temperature with the use of Eqs. (17), (19), (20), (30), (31), (41)-(45),
when we treat
$\alpha_0$, $\beta_0$, $\gamma_0$,
$g_0$, $g_1$, $g_2$,
$T_C$, $\Theta_m$, $\Theta_n$, $p_m$, $p_0$,
and $y$,
as input parameters.
Our strategy for calculating the temperature-
and layer-thickness-dependent MR ratio is as follows:
We first determine the parameters, $\alpha_0$, $\beta_0$ and
$\gamma_0$ to be consistent with the band calculation, and
also $g_0$, $g_1$ and $g_2$ so as to reproduce the $N$ dependence
of the observed, {\it ground-state} parallel GMR.
Then fixing there six parameters thus determined, we calculate the
{\it finite-temperature} GMR with the additional parameters,
$T_C$, $\Theta_m$, $\Theta_n$, $p_m$, $p_n$ and $y$,
which can be properly chosen, as will be discussed
in the model calculations of the next section.


\vspace{1.0cm}
\noindent
{\large\bf III. MODEL CALCULATIONS}

\noindent
A. {\it Fe/Cr Multilyers}

    Gijs {\it et al.}$^{15}$ have observed
both the parallel and perpendicular
GMR for a sample of (3 nm Fe + 1.0 nm Cr) multilayer,
whose results are plotted by circles and squares in Fig.2,
respectively.

    Firstly we consider the case of  $T$ = 4.2 K.
We determine the value of $\gamma_0 = 0.4 $
from the ground-state band calculation of
$\rho_{\uparrow}/\rho_{\downarrow} = 2.3$.$^{22}$
We adopt $\alpha_0 = 7.9$ and $\beta_0 = 1.0$, leading to
$B = 3.38$ (Eq.(42)), which
is consistent  with the value estimated from Eq.(36)  by using
the band parameters such as $\tilde{\varepsilon}^{\rm Fe}$ etc.
We choose the parameters of
$g_0 = 0.045, g_1 = 0.77$, and $g_2 = 3.05$,
such that we have
a good fit to the envelope
of the observed layer-thicknes
($t_N$) dependence of parallel GMR
in (3 nm Fe + $t_N$ Cr) multilayers.$^{11}$

     Next we consider the MR ratio at finite temperatures.
We assume the Curie temperature of the multilayer
of $ T_C = 1000$ K because the thickness of the Fe layers
of the adopted Fe/Cr multilayers$^{11,15}$ is
sufficiently thick to sustain  the Curie temperature of bulk Fe.
The Debye temperatures of Fe and Cr are
assumed to be
$\Theta_m = \Theta_n = 460 $ K.
The phonon parameters, $p_m$ and $p_0$, can be determined as follows:
The total resistivity, $R$, of a {\it pure, bulk} metal is
given from Eqs.(2), (3) and (30), by
\begin{equation}
R(T) \propto  \left\{ \sum_s [(1 + s \: \gamma_0 \: m(T))
(\mu(T)^2 - m(T)^2 + p_m \: Z(T/\Theta_m))]^{-1}
\right\}^{-1},
\end{equation}
from which the ratio of the phonon contribution, $R_p$, to the total
resistivity at $T = T_C$ is given by
$r_p \equiv R_p(T_C)/R(T_C)
= p_m Z(T_C/\Theta_m)/[1 + p_m \: Z(T_C/\Theta_m)]$.
The value of $p_m = 0.69$ is chosen
from the experimental data of $r_p = 0.27$ of bulk Fe
(Fig.1(a)).$^{2}$
We  calculate $p_0$ by
$p_0 = p_m \: A \: (\rho_0/\rho)$ derived from Eq.(38)
with $P_n = P_m$ and $\rho_0/\rho = 0.7$.$^{22}$
The parameters discussed above are summarized in Table 1.
The solid curve in Fig.1(a) expresses
the resistivity, $R(T)$, of bulk Fe
calculated by using Eqs.(43) and (46)
with $\gamma_0 = 0.4$ and $p_m = 0.69$,
which well reproduces the observed data.$^2$

The last parameter $y$, which expresses a concentration
of nonmagnetic atoms in the magnetic layer and which
depends on a sample  employed in a experiment, is
treated as an adjustable parameter.
The parallel and perpendicular GMR of the Fe/Cr multilayer
calculated with $y$ = 0.002, 0.005 and 0.01 are shown  in Fig.2.
Our calculation with $y = 0.005$ well explains
both the $(\Delta R/R)^{\parallel}$
and $(\Delta R/R)^{\perp}$ observed by Gijs {\it et al.}$^{15}$

  In order to study the temperature dependence of the GMR in more detail,
we show in Fig.3,
$\Delta_s \:\: (s = \uparrow, \downarrow)$ as a function of the temperature.
When the temperature is raised, $\Delta_{\uparrow}$
and $\Delta_{\downarrow}$ increase because of
the contributions from spin fluctuations and phonons.
Then the ratio, $\Delta_{\uparrow}/\Delta_{\downarrow}
\: (= \alpha/\beta)$,
changes from 7.9
at $T = 0$ to unity at $T \geq T_C$.
Fig.3 also shows the decomposition
of $\Delta_s$ to various contributions from random potentials
($\Delta_s^r$), spin fluctuations ($\Delta_s^s$)  and
phonons ($\Delta_s^p$).
We note that at $T = T_C$,
$\Delta^s/\Delta = 0.70$, $\Delta^p/\Delta = 0.26$
and $\Delta^s/\Delta^p = 2.65$.
This shows a significant spin-fluctuation contribution,
as suggested from the resistivity data of bulk Fe.$^2$

\vspace{0.5cm}
\noindent
B. {\it Co/Cu Multilayers}

  We have performed a similar calculation to
explain the temperature dependence of parallel and perpendicular GMR
of the (1.2 nm Co + 1.1 nm Cu) multilayer observed
by Gijs {\it et al}.$^{16}$
We adopt $\alpha_0 = 0.7$, $\beta_0 = 8.4$
($\beta_0/\alpha_0 = 14$),$^{16}$ and $\gamma_0 = -0.7$ which
comes from the ground-state band calculation of
$\rho_{\uparrow}/\rho_{\downarrow} \sim 0.15 $
of bulk Co.$^{22}$
We cannot determine the values of $g_0$, $g_1$ and $g_2$ because
the layer-thickness dependence of the parallel
GMR of this series of samples has not been reported.
Then we tentatively adopt $g_0 = 0.13$,
$g_1 = 0.39$ and $g_2 = 0.11$
by scaling the data of similar Co/Cu multilayer$^{23}$
as to reproduce the observed ground-state value of
$(\Delta R/R)^{\parallel} = 0.43$.$^{16}$
The Curie and Debye temperatures  are taken to be
$T_C = 1400 $ K and $\Theta_m = \Theta_n = 445 $ K.
We adopt $p_m  = 1.62$ from the observed ratio of
$r_p = 0.56$ for bulk Co (Fig.1(b)),$^{3}$
and $\rho_0/\rho = 0.3$.$^{22}$
Adopted parameters are shown in Table 1.
The solid curve in Fig.1(b) denotes the temperature-dependent
resistivity of bulk Co calculated by using  Eqs.(43) and (46) with
$\gamma_0 = -0.7$ and $p_m = 1.62$.

The calculated $(\Delta R/R)^{\parallel}$
and $(\Delta R/R)^{\perp}$
of the Co/Cu multilayer are shown in Fig.4, where
$y$ is treated as an  adjustable parameter.
Both the parallel and
perpendicular GMR observed by Gijs {\it et al.}$^{16}$
are fairly well explained by
our calculation with $y = 0.005$.

  Figure 5 expresses the temperature dependence of $\Delta_s$ and
its components, $\Delta_s^r, \Delta_s^s$ and $\Delta_s^p$,
which shows that at $T=T_C$,  $\Delta^s/\Delta = 0.42$,
$\Delta^p/\Delta = 0.53$ and
$\Delta^s/\Delta^p = 0.79$.
Comparing these figures with the corresponding ones of Fe/Cr systems,
we note that  spin-fluctuation contribution
in Co/Cu multilayer is
less significant than in Fe/Cr multilayer.
This fact is expected to be
the main reason why the observed temperature
dependence of the GMR in Co/Cu multilayer is less considerable than
that in Fe/Cr multilayers.

\vspace{1.0cm}
\noindent
{\large\bf IV. CONCLUSION AND DISCUSSION}

    We have discussed the temperature dependence of
the GMR for currents parallel and perpendicular to the multilayer plane.
We have included contributions
from the random exchange potentials,
spin fluctuations  and phonons, which are considered
to be main scattering mechanisms
yielding the resistivity in transition-metal multilayers.
Our model calculations have accounted for the following features
of the observed GMR:$^{11,15,16}$
(1) both the
parallel and perpendicular GMR
are significantly temperature dependent than the (average) layer moment,
(2) $(\Delta R/R)^{\perp}$ is larger than
$(\Delta R/R)^{\parallel}$, (3) the temperature dependence of
$(\Delta R/R)^{\perp}$ is more significant  than  that of
$(\Delta R/R)^{\parallel}$,
and (4) the temperature dependence of GMR in Co/Cu multilayers
is less considerable than that in Fe/Cr multilayers.
The effect of spin fluctuations plays an important role to account
for these three items whereas phonons play a secondary role.
In fact, the items (1)-(3) can be explained without invoking
phonons.$^{10}$

    In our phenomenological analysis, we have assumed that the bulk
scattering is predominant.
On the contrary, when we take into account only
the  interface scattering, the expression for
the GMR is given again by Eqs.(17)-(20) but with $M$  replaced by $I$,
the thickness of the interface, and with $\alpha$ and $\beta$
expressed in terms of the quantities relevant to the interface.
Then, they have ostensibly similar $T$ and $N$  dependence to
those in which only the bulk scattering is included.
It is possible to extend our analysis
taking into account both the interface and bulk scatterings,
although the calculation becomes laborious because it inevitably
needs much number of parameters.
Among many parameters, the most important ones are
$g_0$, $\alpha_0$, $\beta_0$ and $y$;
$(\Delta R/R)^{\parallel}$ generally becomes smaller than
$(\Delta R/R)^{\perp}$ by $g_0$, and the essential feature of
the temperature dependence of the GMR is determined by
the ratio of $\alpha_0/\beta_0$ and $y$.$^{10,12}$

\vspace{1.5cm}
\noindent
{\it Acknowledgment}

    This work is partly supported by a Grant-in-Aid for Scientific
Research on Priority Areas from the Japanese Ministry of Education,
Science and Culture.

\newpage
\noindent{\large\bf  Figure Captions}   \par
\vspace{0.5cm}

\noindent
{\bf Fig. 1}  The temperature dependence of the
observed resistivity (circles)
of (a) {\it bulk}  Fe  (Ref.3) and (b) Co (Ref.4);
the calculated resistivity, $R(T)$, and its  phonon term, $R_p(T)$,
above $T_C$ are shown by solid and dotted curves, respectively,
results being normalized by $R_C = R(T_C)$.
\vspace{0.3cm}

\noindent
{\bf Fig. 2} The temperature dependence of
the parallel ($\parallel$) and perpendicular $\Delta R/R$ ($\perp$)
of (3 nm Fe + 1.0 nm Cr) multilayers.
Dotted, solid and dashed curves denote the calculated results
with $y$ = 0.002, 0.005 and 0.01, respectively;
circles (squares) expressing the observed parallel
(perpendicular) GMR (Ref.15).
\vspace{0.3cm}

\noindent
{\bf Fig. 3} The temperature dependence of $\Delta_s$
of up-spin (solid curves) and down-spin electrons (dashed curves)
calculated with $y = 0.005$ for the (3 nm Fe + 1.0 nm Cr) multilayer.
Also shown are their decomposition to various contributions
from the random exchange potentials ($\Delta_s^r$),
spin fluctuations ($\Delta_s^s$) and phonons ($\Delta_s^p$);
the calculated results being normalized by
$\Delta_C = \Delta_s(T_C)$.
\vspace{0.3cm}

\noindent
{\bf Fig. 4} The temperature dependence of the parallel ($\parallel$) and
perpendicular $\Delta R/R$ ($\perp$)
of (1.2 nm Co + 1.1 nm Cu) multilayers.
Dotted, solid and dashed curves denote the calculated results
with $y$ = 0.002, 0.005 and 0.01, respectively;
circles (squares) expressing the observed parallel
(perpendicular) GMR (Ref.16).
\vspace{0.3cm}

\noindent
{\bf Fig. 5} The temperature dependence of $\Delta_s$
of up-spin (solid curves) and down-spin electrons (dashed curves)
calculated with $y = 0.005$ for the (1.2 nm Co + 1.1 nm Cu) multilayer.
See a caption of Fig.3.

\end{document}